\documentclass[preprint, authoryear]{sigplanconf}

\usepackage{underscore}
\usepackage[hyphens]{url}
\begin{document}


\title{CVC Verilog Compiler -- Fast Complex Language Compilers Can be Simple}

\authorinfo{Steven Meyer}
           {Tachyon Design Automation\\
           Boston, MA 02109, USA}
           {\mbox{smeyer@tachyon-da.com}}
\maketitle
\begin{abstract}
This paper explains how to develop Verilog hardware description language (HDL)
optimized flow graph compiled simulators.
It is claimed that the methods and algorithms described here can be
applied in the development of flow graph compilers for other
complex computer languages.  The method uses the von Neumann computer
architecture (MRAM model) as the best abstract model of computation
and uses comparison and selection of alternative machine code sequences
to utilize modern processor low level parallelism.
By using the anti formalist 
method described here, the fastest available full IEEE 1364 2005 Verilog
HDL standard simulators has been developed. 
The compiler only
required 95,000 lines of C code and two developers.
This paper explains how such a compiled simulator validates the anti-formalism
computer science methodology best expressed by Peter Naur's 
datalogy and provides specific guidelines for applying the method.
Development history from a slow interpreter into
a fast flow graph based machine code compiled simulator is described.
The failure of initial efforts that tried to convert a full 1364 compliant
interpreter into interpreted execution of possibly auto generated virtual
machines is discussed.  
The argument that fast Verilog simulation requires detail removing
abstraction is shown to be incorrect.
Reasons parallel GPU Verilog simulation has not succeeded are given.
\end{abstract}
\section{Introduction}
\par
This paper presents a novel method for implementing fast compilers
for complex computer languages.  The simple organization
and development methods
used to create a fast Verilog hardware description language (HDL)
machine code simulator are described.
In electronic design, systems are expressed as hardware
description language code.  The HDL describes both procedural
parts of electronic systems and declarative gate level parts.
In the case of Verilog, the HDL is a low level language similar to
Pascal \cite{Wirth1971} with added low level parallelism and electronic
gate descriptions. 
The Verilog compiler translates the HDL description into 
machine code for X86 and X86_64 machines running Linux that is executed
simulating electronic system behavior after fabrication.
All electronic design automation (EDA) tools (also physical design
tools not discussed here) run only on enterprise
quality Linux because of large size of EDA system descriptions.
\par
The organization of the compiled simulator can be viewed as a modern analog
of the multi-pass compilers applying problem specific methods used to
develop early compilers \cite{Gries1971} (8-10 for history, 410, 411 and
451-454 for methods).  The early compilers were  developed by scientists
who were mostly trained as physicists. See Budiansky's account
of the role of English physicist Patrick Blackett's department
during WWII that was perhaps the first use of modern algorithmic thinking
\cite{Budiansky2013}.
\par
Computers are now so fast and are equipped with so much
fast random access memory that the various (virtual) passes can be 
executed without storing results of each pass in secondary storage.
The organization in which one unified representation
is continually modified and transformed is similar to the approach
Peter Naur used in developing the Gier Algol compiler \cite{Gries1971} (p. 9).
For example, pass 6 for a compiler run on a machine with only 128,000
words of memory checked types of identifiers and operands (and
updated information) then converted to Polish notation and output
the result for the next code generation pass.  In the simulator
described here, the code generator
does similar checking but converts the source program (really the internal
representation of source because of ample memory) into a flow graph of basic
blocks that contain virtual machine instructions.
\section{Simulator Description}
\par
The Verilog hardware description language is defined by the IEEE 1364-2005
Verilog HDL standard \cite{IEEE2005}.  The standard includes
both register transfer level (RTL) procedural descriptions and
gate level descriptions with rules for annotating accurate delay information
from manufacturing process measurements.
\par
The simulator described here was developed by two people in less
then 10 years.  The ten years included training a young programmer who
started as a college student intern and includes a majority 
of time implementing the elaboration code for the generate feature added
by the 1364 standard committee which is almost incompatible with
instance based display machine register technology required for
very fast Verilog simulation (\textbf{__idp} area).
\par
The simulator consists of about 330,000 lines of C code that is 10\%
or less the size of the competing commercial quality Verilog
full accuracy simulators.  The simulator was first developed as an interpreted
Verilog simulator, then the flow graph based compiler was added.
The Verilog elaborator (parser, fix up and simulation 
preparation phases) is about 100,000 lines (numbers are approximate
because there are common routines and overlap).  Of the 100,000 lines,
20,000 or 20\% are needed for the complicated Verilog 2005 generate feature.
\par
Generate allows compile time variables called parameters to
change HDL variable sizes and design instance hierarchy structure.
This feature creates instance specific constants that must be
treated as variables during simulation.
The other Verilog simulators flatten designs resulting in
much larger memory use for designs with many repeated instances
and makes based addressing difficult.
It is common for an IC to contain millions of instances of a latch
or flip flop macro cell each of which requires storing not just its
per instance state information but also the machine instructions
for every instance when the flattened elaboration method is used.
In the simulator described here machine code to simulate one instance
plus per instance state information and per instance base addressing change
machine instructions only need to be stored.
The development of the Verilog 1364 standard
by committee has continually made Verilog more complicated
(the language reference manual is 590 pages) and has made it more difficult
to implement the simulator's per instance base address register algorithms.
The  display offset algorithm results in a simpler code generator
and faster simulation at the cost of a much more complicated
elaboration phase.
\par
About 50,000 lines are needed to execute interpreted simulation.
The compiler itself is only about 95,000 lines of which 15,000
are executable binary support libraries.
70,000 lines are needed to implement miscellaneous features used
both by the compiler and the interpreter: SDF delay annotation,
four programming language interface APIs (tf_, acc_, vpi_ and dpi_),
a debugger for the interpreter, toggle coverage recording and report
generation, rarely used switch level simulation, plus an expression
evaluation variant called X propagation that implements a more
pessimistic unknown (X state) evaluation algorithm.
\section{Relation to Naur's Datalogy and Anti-Formalism}
\par
The theoretical background behind this simulator's development method
follows Naur's methods.  In the 1990s Peter Naur, one of the founders
of computer science, realized that CS had become formal mathematics separated
from reality.  Naur advocated the importance of programmer specific
computer program development that does not use preconceptions.  The clearest
explanation for Naur's method that was used in developing this compiler
appears in the book \textit{Conversations - Pluralism in Software
Engineering} \cite{Naur2011}.
This books amplifies the program development method Naur described
in his 2005 Turing Award lecture \cite{Naur2007}.
In \cite{Naur2011} page 30, the interviewer asks ``... you basically say
that there are no foundations, there is no such thing as computer science,
and we must not formalize for the sake of formalization alone''.
Naur answers, ``I am not sure I see it this way.  I see these techniques
as tools which are applicable in some cases, but which definitely are
not basic in any sense.''
Naur continues (p. 44) ``The programmer has to realize what these alternatives
are and then choose the one that suits his understanding best.
This has nothing to do with formal proofs.''
Einstein described this as the 20th century split between axiomatics
and reality by saying that ``axiomatics purges mathematics of all
extraneous elements'' which makes it evident ``that mathematics as
such can not predict anything'' about reality \cite{Einstein1921}.
See also \cite{Naur2005} and \cite{Meyer2013}
for more detailed discussion of Naur's anti-formalism.
\par
Current compiler development methodology chooses formal algorithms
over Naur's programmer specific approach that rejects pre-suppositions.  During the development of this compiler, anomalies in mathematical
foundations of logic that current computer science takes as truth beyond
criticism influenced design decisions. 
The first is Paul Finsler's proof that the continuum
hypothesis is true \cite{Finsler1969}.  The proof is only indirectly
related to computer science.
The second example is Juri Hartmanis' proof that P=NP in MRAM (parallel
ram) models \cite{Hartmanis1976}.
This power of MRAM machines contributed to looking for other sources of
parallel speed improvement and motivated using von Neumann
machine architecture targeted language specific recursive descent parsing.
\par
A more direct result of rejecting formalism is the choice
to use the fast Cooper-Kennedy dominator algorithm \cite{Cooper2006}.
This choice and its use in the crucial optimization algorithm define-use
lists is easy once one starts with skepticism toward algorithms that have
been formally proven to be correct.
The Cooper algorithm's ``real'' speed is viewed as
falsification of the axioms used in concrete complexity theory.
\par
The main lesson of this anti-formal development method avoids using
abstraction.
Avoid machine generated tables and compiler phases generated by automatic
generators.
Parse using language specific recursive descend and use the C run time
stack for remembering context.  Parse expressions using simple
operator precedence.  This results in a very fast parser 
but is only possible because assignment operators are not part of expressions
in Verilog.  Assume proofs use axioms that do not apply to reality unless 
specifically determining that the axioms are good.  Finally, following Naur,
look for algorithms that are simple so they can be adopted to the problem
specific aspects of Verilog and the data structures and algorithms
of electronic design simulation.
\section{Importance of von Neumann Observation on Weakness of Formal Machines}
\par
There is a history of compiler development that was closely tied
to the von Neumann computer architecture.  The best model
of computing is the von Neumann machine itself.
Von Neumann was skeptical of models that used
automata in general and understood that Turing Machines (TM) were too
weak to model human program writing expressed in his 1950s criticism
of neural networks and other simple automata.  Von Neumann's thinking is
analyzed in detail in \cite{Aspray1990} and \cite{Meyer2016}.
Von Neumann explicitly justified machine models that consist of unbounded
memory cell size finite number of randomly accessible memory cells.
\par
TMs are too weak a model of computation because they lack random access
and lack the ability to select bits from cells.
In other words, the P=?NP question only exists for TMs not for
von Neumann architecture MRAM machines for which the class P is equal
to the class NP \cite{Hartmanis1976} \cite{Meyer2016}.
For von Neumann machines, guessing (non determinism) is no faster
than enumeration in the polynomial bound sense.
This is expressed in CVC by the use of optimization
algorithms that search not exhaustively but using graph theory properties
combined with human conceptual problem solving.
For decidable problems outside the class NP such as the yes or
no question ``are two regular expressions equivalent?'',
compiled computer languages
allow people to code their understanding into a computer program or
electronic circuit description so that the program can solve
concrete regular expression equivalence problems.
\par
Two concrete examples of traditional compiler development
anti-formalism are William Mckeeman et. al. development of the
XPL compiler system \cite{Mckeeman1971} and development of Bell
Labs C computer language and compiler (also development of Unix)
\cite{Ritchie1993}.  Mckeeman's contribution was understanding
compiler boot strapping.
Dennis Ritchie and Ken Thompson understood that the model they
were developing a compiler and operating system for was the
von Neumann computer itself instead of developing for some abstract
model of computation that was for example attempted in the  
Multics project.  These pioneers of compiler development also
understood the importance of methods that allowed one or two people to
develop large computer programs.
\section{Development History}
\par
The simulator described here was developed in the 1990s as a
Verilog simulator to compete with the original Verilog XL
simulator that used interpreted execution.
At the time Verilog semantics was aimed at interpreted simulation
because simulation properties such as delays could be set at any
time during a simulation run and because a command line debugger was
an integral part of Verilog, i.e. simulations often expected to
read Verilog source from script files at various times during a simulation run.
In the late 1990s Verilog native machine code compilers were introduced
and Verilog semantics was changed to require specification of all
design information and properties at compile time.
See \cite{Thomas2002} for a historical description of the
Verilog HDL.
\par
By 2000 this simulator was no longer speed competitive so it was used as the
digital engine for an analog and mixed signal (AMS) simulator.
The speed problem was not as serious for mixed signal
simulation because analog simulation requires solving differential
equations.  Analog simulation numerical algorithms run orders
of magnitude slower than digital simulation.
\par
After the use in an AMS simulator, the simulator described here needed
to be improved so that it would be speed competitive with compiled
digital Verilog simulators.
The most obvious speed problem involved register transfer level (RTL)
simulation.  RTL simulation is almost the same as normal programming
language execution except Verilog values require at least 2 computer bits
per Verilog bit and the RTL execution must interact with an event driven
scheduler.
\par
The most obvious problem was that a number of \textbf{if} statements in the
interpreter C code were needed to select which interpreter algorithm
to run.  For example, a simple logic and (\&) operator
needs different evaluation C language code sections (usually one procedure)
for scalars (1 bit), narrow vectors (up to 32 or 64 bits), wide vectors
and strength model bit vectors (one byte per bit required).
In addition for all but scalars, both unsigned and signed execution
procedures are required. 
Sign extension for non integral number of word bit vectors
requires significant calculation.  The Verilog standard requires that at least
up to one million bits wide vectors simulate correctly.
It seemed that taking the interpreter evaluation routines and converting
to high level instructions for a virtual machine that could then be
interpreted was a good idea (see for example \cite{Ertl2003}).
Automatically generating a virtual Verilog machine from the simulators
interpreted evaluation code was tried \cite{Ertl2002}.
The development was not too hard, but the resulting execution speed
increased performance by only a small amount.
Although the \textbf{if} statement overhead
was removed, extra overhead to decode and execute the interpreter
virtual instructions nullified most of the gains.
\subsection{Concrete algorithms but not organization from Morgan's optimizing compiler book}
\par
It was realized that the instruction level parallelism and branch prediction
provided by modern microprocessors was required for fast
simulation.
Development of a full code generator began.
It was next attempted to implement the concrete step by step method
in Robert Morgan's book on building an optimizing compiler \cite{Morgan1998}.
The book describes the method used in building the very good Digital
Equipment Corporation Alpha microprocessor compilers.
\par
This simulator does not use the code generator organization
from \cite{Morgan1998} (section 2.1, 21-26).
Morgan advocates a basically breadth
first filter down approach with a chain of transformations each of
which has a different data representation.  Morgan
writes: ``Each phase has a simple interface'' that can be tested in
isolation.  Also, ``No component of the compiler can use information
about how another component of the compiler is implemented'' (p. 21).
\par
Instead, modified versions of the very good algorithms
spread throughout the Morgan book are implemented.
During development of the code generator, new ideas were
compared against the concrete Morgan book approach to make sure they were no
worse.  The Morgan book algorithms are especially useful because
exceptions are discussed. For example allowing non static
single assignment (SSA) form lists that violate the
rule that each variable is assigned to only once or allowing 
define-use chains sometimes to not have exactly one element (p. 142).
\par
Morgan's idea that virtual instructions should be as
close to machine instructions as possible is used (Figure 2.2, p. 24).  The
one exception is that Verilog requires very complicated mostly
boiler plate prologue and epilogue instruction
sequences that is just one virtual instruction.
\par
Code generator steps use one master representation accessible from the
interpreted data base.  Both flow graphs and temp names
(unbounded number of virtual registers that in Verilog are often wide and
require two bits to represent 4 values) are accessed in numerous ways.
Basic blocks are accessible directly from interpreter execution form that then
points to flow graphs.  Flow graphs especially for net change propagation
operators are accessed from indexed tables and AVL trees (see \textbf{igen.h}
in simulator source).  The machine code compiler adds information to the
previous interpreted simulator data structures.
\par
A compiler organization that combines all information into one
master data base is best.  Any code generation phase can use any
of the information that is accessed either directly from the interpreter
execution net list, from indices (sometimes indexed tables and sometimes
trees), from code generation basic blocks or machine code routines.
The idea is to continue to improve the global data base during code generator
development and to continually add more information to all of the various
parts of the one unified representation.  For example, flow graph
building algorithm insights were used to improve design elaboration
data structures and interpreter execution data structures.
\par
This unified data base where each part is kept consistent with other
parts is the key to simplicity and code quality.  For Verilog,
because there are so many different types of operations from
procedural RTL to declarative gate level to load and driver propagation,
depth first code generation is better.  Morgan's type of low level
machine instructions (p. 24) are generated with some optimization by
expanding constructs all the way down to something close to the final
virtual instruction sequences when the flow graphs virtual instruction
sequences are built.  Later mapping to machine instructions except
for X86 fixed registers is straight forward.  This approach may be
Verilog specific because Verilog allows values to be read and written
from anywhere in a design.  Cross module references are allowed and
a programming language interface (PLI) can run in any Verilog thread. 
There is effectively no usable context information in Verilog.
\par
For example, once some C code for the unified data structures of the flow
graph and basic block mechanism were written, the Morgan book detailed
algorithms on define-use lists and importance of SSA
(12.5.1 p. 291 and 7.1, p. 142) could be applied.
The heuristic is used to generate (top down depth first) as many
temporaries as possible.  Then SSA problems are fixed when needed or even
allowed to violate SSA form that is recognized by the machine code
expander..
The crucial data structure used for optimization is the
define-use lists.  Morgan suggests the idea
of allowing non SSA instructions and temporaries (p. 142).  Extra
temporaries can then be eliminated during optimization passes through
all flow graphs and virtual instruction lists.
\subsection{Compiler file organization}
\par
Basic block creation and virtual instruction generation C code is
in \textbf{v_bbgen}'s C files.  Define-use list and
other flow graph elaboration code are in the
\textbf{v_bbopt.c} file. 
The \textbf{v_regasn.c} file assigns machine registers to the unbounded number
of temps.  The \textbf{v_cvcms.c} and \textbf{v_cvcrt.c} files
plus the \textbf{v_asmlnk.c} file contain support C procedures
plus code to generate
the GNU AS assembly output, run \textbf{gas} and link the final output
executable called \textbf{cvcsim}.
The \textbf{v_aslib.c} file contains wrapper
C procedures that are called from the generated assembly but whose function
is usually to call an interpreter execution procedure.
By using wrappers, early
versions of the compiler could compile almost all of Verilog but simulation
was not yet fast because execution used wrappers that executed the
slow interpreter code.
\section{What is Verilog}
\par
Verilog is a Pascal like language \cite{Wirth1971} for the description
of electronic hardware.  All variables are static because there are
no implicit stacks in hardware.  Verilog is a combination of normal behavioral
programming with parallelism, execution of hardware described at the
RTL level and low level primitive declarative
gates and flip flops.  A common electronic design method codes circuit
descriptions in RTL then runs a program to synthesize the RTL into
gates also coded in the Verilog language.  Verilog is used to simulate
(predict behavior when an IC is fabricated) both the RTL and the synthesized
gates with accurate timing.  See \cite{Thomas2002} for a description of
the Verilog HDL.  See \cite{Sutherland2006}, pp. 401-413 for a history 
of the Verilog HDL.  See \cite{Allen2002}, pp. 619-622 for a description
of Verilog from the viewpoint of optimizing compiler development.
\section{Why Verilog Simulations Run Slowly Compared to Computer Programs}
\par
First, Verilog RTL values require 2 bits (4 values) for every hardware bit.
Gate level accurate delay simulation also requires bus values that
allow 127 different values and driving strengths.  A simple logic
operation requires at least 3 or 4 instructions (not counting loads and
stores).  Second, hardware registers are almost always wider than
the native von Neumann machine register width because hardware design involves
modeling the next generation electronics.  This requires evaluating
multiple machine words for each operation.  Third, Verilog requires  
event driven simulation.  When a delay or event control (@(clk3) say)
is executed, the simulation must schedule a new event in an event
queue and suspend the current execution thread to be restarted later.
A even slower process occurs when a value is changed.  All variables
that are referenced in right hand side expressions driving the
left hand side value must be evaluated and new assignments made.
These effected left hand side variables are called
loads.  Also, when a wire is evaluated, it may be necessary to evaluate
multiple drivers and determine which is the strongest.  See \cite{Meyer1988}
for a data structure that allows implementation of efficient load propagation
and driver competition algorithms.
\section{Performance}
\par
The simulator described here is arguably the fastest full accuracy 1364
Verilog simulator.
Evidence here is anecdotal from users because the other commercial
simulators are licensed with no public benchmarking allowed clauses.
One anonymous published comparison of this simulator with other
compiled commercial simulators for RTL but not gate
level simulation speed \cite{Semiwiki2011} showed 2 to 5 times
faster simulation by this simulator for two discussed designs.
\par
Other anecdotal evidence is that
optimization is so good that
in one case full accuracy 1364 Verilog simulation is almost as fast as
cycle based evaluation using a detail removing cycle based
open source Verilog evaluator called Verilator for a Motorola 68k
processor model.
Such fast simulation is possible because the model really only
requires 2 state not 4 state variables.
\subsection{Comparison with Open Source Icarus Verilog}
\par
There is another open source Verilog simulator Icarus
Verilog called \textbf{iverilog} that implements most
IEEE 1364 Verilog RTL features.
It is difficult to use for benchmarking because it does not
implement Verilog source macro cell library options
(called \textbf{-v and -y} libraries).
A sixty four bit version of the simulator described here is usually
35 to 45 times faster than Icarus Verilog release
called \textbf{10.0 stable} made with the 
default options (-g -O) run on an Intel Core i5-5200U 2.2 GHz
low power processor running Linux Centos 6.7.
\par
For one SHA1 check sum circuit, this simulator is 111 times faster.
The SHA1 model is almost just a Pascal program so again
the +2state option can be used.
The two state option works by keeping
the X and Z values (called B part words) around but simulation
only needs to initialize the words to zero once.
If a design simulation really requires X and Z values,
simulations run with this option will be incorrect.
Flow graph optimization normally removes B part basic blocks from output
machine code when X or Z values are impossible.
The two state option allows generating simpler flow graphs that
allows evaluation of the values for A (0 or 1) words to
be optimized even more.
\section{Simple Design Method for Complex Language Compilers}
\subsection{Develop an interpreter using language specific simple concrete methods}  
\par
This simulator was developed at the same time the Verilog IEEE 1364
Verilog standard was defined and being continually changed.
It is very difficult to simplify or shorten compiler development
duration when a computer language is also under development.
Some helpful ideas are:
\subsection{Use simple organization}
\par
One centralized include file is used
by every C source file with defined and used prototypes at the top of each
source file.
This organization makes it easy to eliminate occurrences of more than
one routine for the same basic function
(wide vector sign extension for example).
Also, it is important to be able to make a change in only one place
when Verilog changes or when better simulation algorithms are implemented.
It is also usually not obvious when a new Verilog feature
is first introduced what part of simulation is the same as some pre existing
feature.
\subsection{Use only one internal organization (net list data structure for Verilog)}
\par
The first pass scans and parses Verilog source into normal module lists,
statement and gate lists, expression trees and linked symbol tables.
Then the next set of fix up procedures is used to fill the same date
structure with more information which includes possibly totally changing
the instance tree hierarchy.  Then the next elaboration phase fills in
more simulation preparation information and allocates variable and state
memory just before beginning interpreted simulation.
This organization allows easily
moving processing steps forward and backwards when the Verilog language
changes.  The compiler is run just before interpreted simulation would
have been executed.  It outputs a binary that is then run for compiled
simulation.
\subsection{Avoid generators, grammars and tables}
\par
The simplest and most powerful scanning method uses a giant
case statement.  The simplest and most powerful parsing method
uses language specific recursive descent because normal push
down stack based parser tables use very weak push down automata,
even weaker than TMs.
This is especially important
in complicated languages such as Verilog where the scanner needs parser
information and the parser needs scanner information.  The 1364 standard
committee does not worry about constraining the language
so that it can be coded as a context free grammar.
Verilog assignments are separate from expressions so 
simple and fast operator precedence parsing can be used
\cite{Gries1971} section 6.1, 122-132. 
This simulator can elaborate 5 million line designs is less than 15 or 20
seconds on a modern fast CPU.
Fast elaboration speeds up development of code generator algorithms,
allows faster compiler debugging and assists in finding faster
machine code instructions patterns for the output machine language
program.
\subsection{For complex languages first develop an interpreter}
\par
The first step in fast compiled Verilog development requires developing
a 1364 standard compliant interpreted simulator so that compiler machine
code execution can be regressed against the interpreter standard.
Once simulation is running, many
speed improvements will become obvious (made visible)
without needing to wait for a long design phase
before experimental data is available.
\subsection{Add interpreter wrapper capability}
\par
There are \textbf{I_CALL_ASLPROC} and \textbf{I_CALL_ASLFUNC}
virtual instructions that work with the normal unbounded register temps and
define-use predominator algorithms, i.e. interface is no different from a
low level virtual machine instruction.
This allows running simulations with only some constructs compiled.
During initial development, only narrow (less than 32 or 64 bit)
variables were compiled.  All wider expressions and assignments were
evaluated with wrappers.  Then step by step wrappers were
replaced by low level machine instructions.
Some very complicated Verilog algorithms such as multiple path
conditional delay selection and switch level simulation are still
just wrappers.
\subsection{Make writing virtual instruction generation same as interpreter execution}
\par
For example, Verilog binary operator evaluation starts with a wrapper,
then is replaced by a version of the same procedure with calls to 
\textbf{__gen_tn} to create temporary registers, and \textbf{__start_bblk} to
replace \textbf{if} statements (see \textbf{eval_binary}
procedures in \textbf{v_ex2.c} file near line 6600) for which
the code generation procedure is in file \textbf{v_bbgen.c}.
This allows code generation to mostly involve replacing C program
evaluation statements with the corresponding 
temp register, basic block  and virtual operation
generation procedure calls.
Code generation for more complicated simulation operations such as
scheduled event processing can be simplified this way also.
\subsection{Modify Morgan's dominator-based optimization algorithms}
\par
The book \textit{Building an Optimizing Compiler} by Robert Morgan
contains a concrete simplified method for optimizing flow graphs
and computing define-use data structures.  Start with that and simplify
even more and modify to fit your complex language.
This is the hard very important part of developing a fast compiler.
Once good flow graph optimizations are implemented, the register allocator
becomes much easier and better.
See \textbf{__optimize_1mod_flowgraphs} at the beginning of file
\textbf{v_bbopt.c} for the code that implements this.
\subsection{Use low level virtual instructions close to machine instructions}
\par
Virtual instructions are defined and emitted that are mapped into
machine instructions.  It is best to generate low level virtual instructions
because the code generation human coder for the particular Verilog feature
can craft good instructions sequences.  Verilog operations at minimum
require operating on an A part (one or more computer
RAM words) for the 0 and 1 value and a B part for the
X and Z value so even simple operations require multiple machine instructions
and probably two by two vector cross product evaluation.
\par
The copy operation is an exception especially for Verilog because
much simulation is copying data that can be very wide.  Initial code
generation inserts a complicated virtual \textbf{I_COPY} instruction
whenever there is a possibility of a need for a copy.  Then during
mapping from the \textbf{I_COPY} virtual instruction to machine instructions,
the copy usually can be removed without needing to emit anything.
\subsection{Avoid extra representations such as tuples and breadth first transformations}
\par
One of the best simplifying and simulation speed increasing methods
is depth first code generation.  This method is intentionally the opposite
of what \cite{Morgan1998}, page 212 recommends.  Morgan recommends breath
first code generation with successive transformations to lower levels.
It is much simpler to use depth first instruction generation of almost machine
instructions because it allows the compiler writer to control machine
code sequences and reduces number of lines of compiler code.
\subsection{Use experimentation to find fast CPU instruction parallelism code sequences}
\par
Once a working code generator was written and debugged, the best
speed improvement method was to set up shell scripts that would run
two different versions of the compiler on a speed test regression suite
and compare results.  The low level machine instruction sequence from the best
result would then be used.
The developers did not have access to the X86_64 multi-issue pipeline
optimization rules documentation so they needed to run experiments.
Maybe the experimental
approach is better in general because the rules are so complex.
This was the most important simulation execution speed up idea.
The other was compiling the change propagation and scheduled
event processing code into flow graphs.
\section{Allen and Kennedy Hardware Simulation as Abstraction Contradicted}
\par
In the book \textit{Optimizing Compilers for Modern Architectures,}
Randy Allen and Ken Kennedy argue that the task of optimization of
hardware descriptions abstracts
hardware descriptions to a less detailed level \cite{Allen2002}.
Allen and Kennedy write ``Another way of saying this is that simulation
speed is related to the level of abstraction of the simulated design more
than anything else -- the higher the level, the faster the simulation''
(p. 624). For Verilog this claim is wrong.  Low level exact Verilog
and machine code detail modeling results
in faster simulation because it allows maximum utilization of low level
instruction parallelism that is built into modern microprocessors. 
The optimizations include:  processing
as wide a bit vector chunk at a time as possible, 
output instruction sequences to keep instruction pre fetch
queues and pipe lines
full and output instruction sequences that work well with 
microprocessor branch prediction algorithms.
Basic blocks in Verilog are small but very numerous because separate
A parts and B parts require conditionals.
\par
This compiler development project shows that a not very complicated code
generator (less than 100,000 lines of C code) can produce good code
using the experimentation method of running speed regression test suites with
different version of to be emitted instruction sequences and choosing the
fastest.
The standard way
(almost but not quite required by the IEEE 1364 standard) to store
Verilog 4 value logic and wire values stores values as two separate
machine word areas.  The B part selects unknown
values: X (unknown) and Z (high
impedance) values.  If the B part is zero, the A part selects 0 or 1
values.  Bit and part select operations need to be optimized as 
fast load, shift and mask operations that use mostly boiler plate
instruction sequences.
Four value logic operations can also
be generated as combinations of machine logic instructions combining
the A and B parts.  Although, complex logic operations are sometimes better
implemented using separated A and B parts or even as table look ups.
\par
The remainder of this section criticizes specific ideas for
implementing simulation abstraction from the Allen and Kennedy book.
\subsection{Inlining modules, p. 624}
\par
Expanding modules inline is called a fundamental optimization.
Incorrect because it is much better to use the normal programming language
optimization of not inlining but separating machine code from state
and variable data that are accessed using per instance based addressing
using traditional display technology \cite{Gries1971} (pp. 172-175).
Not only is code smaller but separating state from model description
makes many optimizations visible especially ideas for pre-compiling
event scheduling and processing.
\subsection{HDL level execution ordering, p. 626}
\par
Allen and Kennedy argue for HDL statement reordering optimizations.
Far more important is machine instruction reordering to maximize low
level microprocessor instruction parallelism.
Part of the reason for this is that although
Verilog has fork-join constructs, they are rarely used.  Parallelism
in Verilog comes from a very large number of different always block usually
triggered with an event control (for example \textbf{always @(clk)}).
\subsection{Dynamic versus static scheduling, p. 627}
\par
This section discusses oblivious evaluation.  Instead of the normal
Verilog algorithm that ``dynamically tracks changes and propagates
those changes'', the alternative blindly evaluates without any
change propagation overhead.  The oblivious method can not work
for Verilog because it is common to have thousands of tick gaps between
edges that have very high event rates.  Static scheduling
and compiling events after change recording into pre-compiled
flow graphs to the extent possible is more important.
\subsection{Fusing Always Blocks, p. 628}
\par
Here Allen and Kennedy are correct.  Fusing always blocks makes a huge
speed improvement.
\subsection{Vectorizing Always Blocks, p. 632}
\par
The idea is that Verilog code generation optimization
should ``rederive the higher-level abstraction that was the original
intent''.  The problem with this is that the decision to code scalarized
or vectored is best left to the designer (HDL generation program).
Vectorizing is not always better because if
only a few bits change per clock cycle
in a wide vector, it is better to simulate the scalarized individual bits.
There is no way to determine bit percentage switching frequency
for a given simulation a priori.  Change detection of vector
selects is fast because
it only requires a mask, shift and xor then branch, but propagation of
the changes can be expensive if the selected part select bits used in right
hand side expressions do not exactly match.  It is much better not
to change the Verilog HDL code, but to find fast instruction sequences.
\subsection{Two-State versus Four-State Logic, p. 637}
\par
Two state simulation is good when it is possible.  The problem is that
the reason for hardware simulation is to find mistakes that will result
in unknown X (IC state that will be sometimes 1 and sometimes 0)
states in the fabricated integrated circuit.
Most real HDL designs do not allow two-state simulation, but if possible
simulation is much faster because evaluations can directly use
hardware machine instructions.
Most four state evaluations are similar to evaluating two by two vector
cross products.
This simulator's low level design feature treats temporaries even for
four state values that require an A part word memory area and a B part
word memory area as one temp.  When an expression evaluation
can be executed as two state, the B part instruction sequences are just
not emitted or optimized away for expressions containing non two state
elements during flow graph simplification.
Verilog variables can be declared as having a two state type.
\subsection{Rewriting Block Conditions, p. 637}
\par
Allen and Kennedy advocate changing trigger conditions on Verilog
blocks by abstracting and guessing user intent to 
eliminate the need to propagate change operators that in Verilog
are usually event controls on always blocks. Instead of trying
to rewrite the Verilog source to higher level, it is better to
compile event queue propagation by generating flow graphs because most
event controls are simple.
Change operator occurrence scheduling and wake up event processing
should also be compiled into flow graphs.
State data should be first coded into the per instance (\textbf{_idp})
based display area (see procedure
\textbf{alloc_fill_ctevtab_idp_map_els} near
line 3820 of file \textbf{v_prp.c}).  Then separate flow graphs are
coded to schedule the
changes and to propagate the changes.
The generated flow graphs are optimized
the normal way so most change operators require just a few instructions
to schedule and a few instructions when the flow graph is jumped to from
the event queue processing code.  The scheduling part for delay controls
(\textbf{@(clk)} say) flow graph generation procedure
is in the \textbf{__gen_dce_schd_tev} routine near line 10400 in file \textbf{v_bbgen3.c}.
\section{GPU or Other Multi-Core Parallelism Does Not Work for Verilog}
\par
There have been many projects that attempt to use specialized computer hardware
rather than optimizing flow graph compilers to speed up Verilog simulation.
The reason for this is that Verilog simulations are a significant consumer of
electronic company compute cycles with sometimes entire server farms
dedicated to running only Verilog.  Verilog compiled executable speed is of
crucial importance and has huge economic value.
At least so far efforts to speed up HDL simulation either with special
purpose hardware or GPUs have failed.  It is possible to build special
purpose hardware that will run unoptimized Verilog 10 times faster.
The problem is that a good flow graph based optimizing compiler run
on a modern fast multi-issue microprocessor can be 10 times or more faster
than the naive unoptimized computer software algorithms the special
hardware is compared to.
\par
Hardware emulation uses a different approach in which a design is converted
to gate level and then fabricated into FPGAs (``layed out''). 
Emulated designs run much faster than software compiled
simulation so they are good for early software
development, but do not ``simulate'' a design in sufficient detail for
hardware debugging.
\par
The reason parallel Verilog simulation has not succeeded for
most large system models is that Verilog HDL basic blocks
tend to by very small with a high proportion of jumps and synchronizations.
Small basic blocks parallelize efficiently on multi-issue CPUs but
have too much synchronization overhead for coarser parallelism.
\par
In the compiler described here, parallelism
is used in one place.  Value change dump file output (FST format is smallest
and fastest) that is used by wave form viewers (software oscilloscopes)
to debug hardware may be run on up to 2 additional cores to encode value
changes.
This use of parallelism can improve simulation times for simulations that
generate huge value change dump files by a factor of two.
Very large value change files are also written because
they can be input into back end physical design tools that analyze the
value changes to optimize IC power usage.
\par
One area where parallel Verilog simulation has worked to some
extent is providing tools that guess and advise users
how to partition designs among multiple X86_64 cores.
Simulation can then
be run in parallel on multiple cores because the assignments to cores results
in minimal communication overhead.
Parallel Verilog simulation on GPUs has not succeeded.
\section{Conclusions}
\par
The general area of how to represent digital electronic circuits is
currently rather unsettled.  Many designers would prefer to write
programming language (C usually) code and have a program automatically
convert the code into a hardware design that is then represented
in Verilog (called high level synthesis).
\par
However, If a design really can be coded as a computer
program and implemented
using FPGA integrated circuit technology, it may be cheaper
and better in speed, cost and power to implement the function
in software as a highly optimized compiled low level program
on off the shelf multicore SoCs containing
conventional CPUs, signal processing CPUs, GPUs and other types of CPUs.
Another way of expressing this observation is that if digital
circuits can be expressed using only two state logic with X and Z cross
products ignored, implementation as computer programs may be better. 
\par
In my view there is a need for a simple Verilog like HDL that
is intended to be generated by computer programs - maybe V-- --.
Verilog elaboration and use of constant parameters that are sometimes
not really constant at run time makes sense when HDL designs are
coded by hand because it makes Verilog easier to write,
but not when Verilog is machine generated.  If such
a language eliminated non locality such as cross module references
and force-release statements (originally intended to allow coding
reset buttons), much faster simulation might be possible.  Such
simulation could use some type of graph theory connectivity
compilation.
\section{Acknowledgments}
Andrew Vanvick(AIV in source) was the other CVC developer.
\bibliographystyle{abbrvnat}
\bibliography{cc-fastcomp}
\end{document}